\documentclass[%
 reprint,
 amsmath,amssymb,
 aps,
]{revtex4-1}

\usepackage{graphicx}
\usepackage{epstopdf}
\usepackage{dcolumn}
\usepackage{bm}
\usepackage{csquotes}
\usepackage{amsmath}

\begin{document}

\preprint{APS/123-QED}

\title{Magnetocaloric effect and scaling analysis in superspin glass Co/Au nanoparticles}

\author{Pavol Hrubov\v{c}\'{a}k}
 \author{Adriana Zele\v{n}\'{a}kov\'{a}}%
  \email{adriana.zelenakova@upjs.sk}
\affiliation{Department of Solid State Physics, P. J. \v{S}af\'{a}rik University, Park Angelinum 9, Ko\v{s}ice, Slovakia}

\author{Vladim\'{\i}r Zele\v{n}\'{a}k}
 \affiliation{
 Department of Inorganic Chemistry, P. J. \v{S}af\'{a}rik University, Moyzesova 11, Ko\v{s}ice, Slovakia}

\author{Jozef Kov\'{a}\v{c}}
\affiliation{Institute of Experimental Physics, Slovak Academy of Sciences, Watsonova 41, Ko\v{s}ice, Slovakia}

\author{Victorino Franco}
\affiliation{Department of Condensed Matter Physics, ICMSE-CSIC, Sevilla University, P.O. Box 1065, 41080 Sevilla, Spain}

\date{\today}

\begin{abstract}

Bimetallic Co/Au (core/shell) nanoparticle system exhibiting transition form superparamagnetic (SPM) to superspin glass state (SSG) at low temperatures has been investigated with respect to its magnetocaloric properties. Peaks of magnetic entropy change vs. temperature dependences ($\Delta S_M(T)$) have been observed at a temperature in the vicinity of SPM to SSG transition. Despite of significant thermal hysteresis of magnetization at low applied fields a master curve has been constructed from the $\Delta S_M(T)$ dependences employing scaling analysis. That confirms the investigated transition is of the second order type. Different values of critical exponents calculated from field dependence of relevant magnetocaloric response parameters in the proximity of critical temperature have been found for low and high applied fields. Moreover, the exponents exhibited significant disagreement with theoretical values characteristic of mean field theory as well as Ising models indicating distinct behavior of the SSG system at critical temperature.

\begin{description}

\item[PACS numbers]
75.75.Fk, 89.75.Da, 64.60.Ht, 65.40.gd, 75.30.Sg
\end{description}
\end{abstract}

\maketitle

\section{\label{sec:level1}Introduction}

Despite of intensive research on nanostructures driven mainly by the high application potential of these kind of materials in a variety of disciplines like medicine \cite{1,2}, technics \cite{3} or industry \cite{4,5}, only a few works devoted to examination of magnetic nanoparticles (NPs) with respect to their magnetocaloric properties have been reported so far \cite{6,7,8,9}. The magnetocaloric effect (MCE) refers to conversion of magnetic energy of a magnetic substance into thermal energy under applying or removing an external magnetic field \cite{10}. Since the effect is known over a century, it has been the basis for several magnetic refrigeration techniques and a number of feasible refrigerant materials have already been designed \cite{11}. In the process of tailoring the qualities of potential refrigerants, magnetic nanoparticles have been found very pioneering. The main advantage of nanoparticles over the bulk materials stems in wider options for tuning the characteristics of the refrigerant material. Since the change of working temperature, refrigeration capacity or phase transition character in bulk materials can be induced almost exclusively by chemical composition, in the case of NPs size, shape, capping layer or dilution of nanoparticle system affect those features. Monte Carlo simulations carried out by Baldomir et al. \cite{12} suggest that for a given sample concentration of superparamagnetic NPs there exists a particle size that provides larger entropy increase, and reciprocally for a given particle volume there exists a sample concentration allowing to produce larger entropy change. Significantly different behavior was documented by Phan et al. \cite{13} in the bulk and nanostructured (50 nm and 35 nm) gadolinium iron garnets (Gd$_{3}$Fe$_{5}$O$_{12}$). While the peak of magnetic entropy change ($\Delta S_M(T)$) in bulk sample was found almost independent of the applied magnetic field, the ones corresponding to nanostructured samples tended to shift towards lower temperatures as the applied magnetic field was increased. Moreover, an increase in the magnitude of $\Delta S_M$ was documented with diminishing particle size. The authors ascribed this feature to surface spin disorder in nanostructured samples. Hueso et al. \cite{9} studied MCE properties of La$_{2/3}$Ca$_{1/3}$MnO$_3$ nanoparticles and found out that the first-order magnetic transition can be minimized and eventually evolves towards a second-order one as the grain size diminishes. Thus, we are able to change the nature of an intrinsic property of a material by changing the particle size. The investigation of numerous bulk systems exhibiting MCE by means of scaling analysis shown that in the case of the materials undergoing the second order phase transition (SOPT) an universal curve can be constructed from the magnetic entropy change vs. temperature dependences after proper rescaling \cite{14, 15}. Therefore, the collapse of $\Delta S_M(T)$ curves can be regarded as the relevant indicator of SOPT in bulk crystalline and amorphous alloys \cite{16,17, 18, 19}. Franco et al. \cite{20} extended this kind of analysis to monodispersed core/shell Co, Co/Ag and Ni/Ag nanoparticle systems. They utilized the field dependence of the magnetic entropy change peak for the surface spin freezing transition study and concluded that although the magnitude of the peak entropy change and position of the peak can be tuned by changing the composition and nature (metallic or organic) of the shell and surfactant layers, the characteristics of the spin freezing transition are not altered.

Apparently, there is a need for extended investigation of many interesting phenomena connected with magnetic NPs. Hence, we decided to prepare a monodisperse core/shell Co/Au nanoparticle system and study its magnetocaloric properties. As it was demonstrated elsewhere \cite{21}, this system exhibits  features of non-equilibrium dynamics induced by strong inter-particle interactions what leads to formation of superspin glass state. This work is the first attempt to apply scaling analysis of $\Delta S_M(T)$ \cite{14} developed for bulk materials on nanoparticle system exhibiting superspin glass state in order to examine the characteristics of the phase transition. Since complex study on MCE characteristics of similar system has not been reported yet, we put forward our results of magnetic entropy change, field dependence of relevant MCE parameters, scaling analysis and critical exponents obtained by processing the experimental data.

\section{EXPERIMENTAL}

\begin{figure}
	\graphicspath{ {figures/} }
	\includegraphics[scale = 0.45]{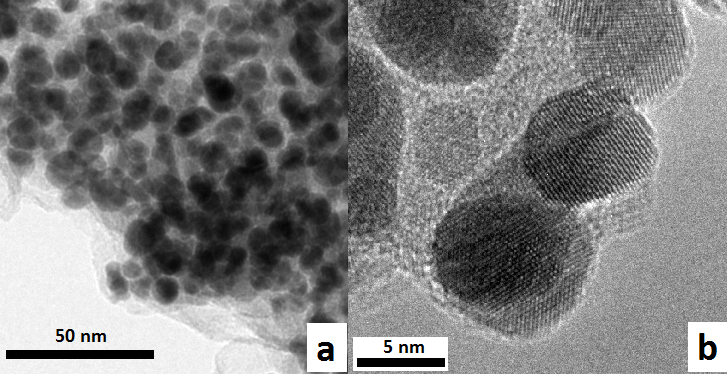}
	\caption{ TEM micrographs of (a) Co/Au nanoparticle asembly and  (b) detail of selected area.}
	\label{figure1}
\end{figure}

Examined Co/Au nanoparticles were synthesised employing the reverse micelle method. Utilized chemicals, process of preparation, particles' structural analysis via X-ray diffraction and high resolution transmition electron microscopy along with histogram of particles' size distribution can be found elsewhere \cite{ 21}. Monodisperse core/shell nanoparticle system (denoted as Co/Au1 in ref. \cite{21}) exhibited average particle size 7 nm with average particle's magnetic moment 86 $\mu_{B}$ corresponding to a cobalt core with diameter of 5 nm. The analysis of AC magnetic susceptibility data revealed the presence of non-equilibrium dynamics and freezing of the nanoparticles' magnetic moments into the superspin glass state at $T_{SSG} \sim 7$ K. The morphology of the investigated system is displayed in Fig. \ref{figure1}.

Magnetic properties of the system were examined by SQUID-based magnetometer (Quantum Design MPMS 5XL). The sample with 18 mg mass was encapsulated in a gelatine capsule and placed into a plastic holder. The signal of the empty capsule and holder was measured and substracted from experimental data. Isothermal magnetization curves were obtained using the following standard protocol which is commonly being used for $M(H)$ data collection \cite{22, 23}. The sample was cooled down in zero magnetic field to 20.8 K and subsequently isothermal magnetization curves were recorded in a temperature range from 20.8 to 1.8 K with a step of 0.5 K in applied fields up to 5 T. Zero-field-cooling ($M^{ZFC}$) and field-cooling ($M^{FC}$) measurements were carried out in temperature range 1.8 - 30 K in applied fields up to 1 T. MCE was evaluated by means of magnetic entropy change vs. temperature dependences calculated from $M^{ZFC}$, $M^{FC}$ and isothermal magnetization data for low and high applied fields, respectively.

\section{Results and discussion}

\subsection{DC magnetic measurements}

\begin{figure}
	\graphicspath{ {figures/} }
	\includegraphics[scale = 0.37]{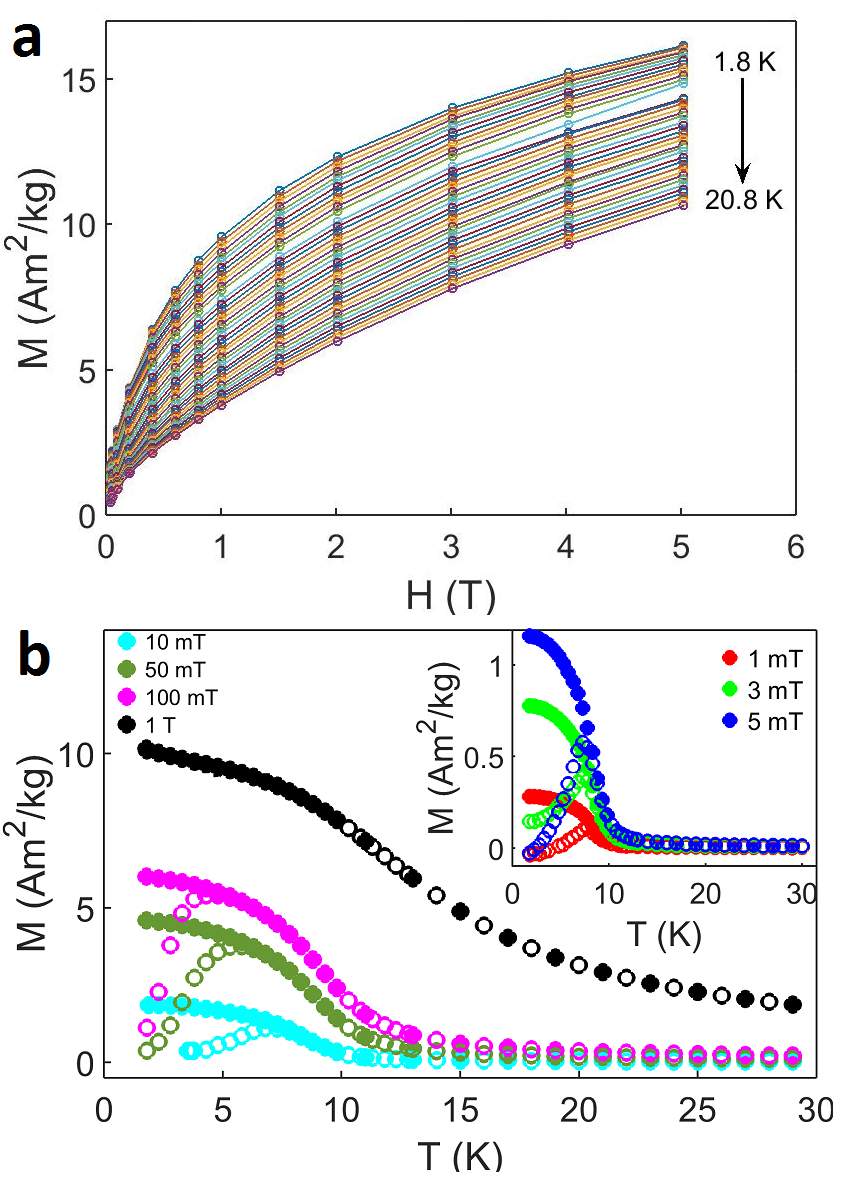}
	\caption{(a) Isothermal magnetization data of investigated Co/Au nanoparticle system recorded with the step of 0.5 K in applied fields up to 5 T. (b) $M^{ZFC}$ (open symbols) and $M^{FC}$ (filled symbols) magnetization vs. temperature dependence of the Co/Au system obtained at constant applied magnetic fields up to 1 T.}
	\label{figure2}
\end{figure}

Isothermal magnetization data recorded by means of standard protocol described in section II, Fig. \ref{figure2} (a), indicate expected dependence of investigated Co/Au nanoparticle system magnetization on applied field magnitude. For each fixed temperature, the magnetization rises with increasing applied field, and for the corresponding applied field magnitude, it decreases with diminishing temperature. 
Fig. \ref{figure2} (b) represents isofield magnetization data, where $M^{ZFC}$ curves exhibit field dependent single maximum at peak temperature ($T_p$) which shifts towards lower temperatures with the applied field increment up to 100 mT. This feature is typical of superparamagnetic NPs and according to Tadi\'{c} \cite{24,25}, the peak temperature, $T_p$, can be attributed to experimental blocking temperature of the system at particular field. Decreasing the temperature below $T_p$ $M^{ZFC}$ and $M^{FC}$ curves start to bifurcate and significant thermal hysteresis becomes evident.The flat character of the $M^{FC}$ curve with tendency to saturate as it approaches the lowest temperatures is one of the characteristic features of superspin glass systems with nonequilibrium dynamics, where strong interparticle interactions are dominant \cite{26}. On the other hand, no thermal hysteresis was observed for $M^{ZFC}$/$M^{FC}$ curves obtained at $\mu_{0}H = 1$ T. These curves merge in the whole temperature span, suggesting that applied field was strong enough to overcome mutual interactions between particles and aligned the particles' magnetic moments even at the lowest temperatures.

In order to examine magnetocaloric properties of the system via magnetic entropy change calculation, isothermal magnetization data obtained by means of standard protocol \cite{22, 23} are commonly being used in the majority of cases. However, it has been shown \cite{22, 27, 28}, that in the systems undergoing first order phase transition where thermal hysteresis is present, the application of Maxwell relation to the $M(H)$ curves could possibly give unphysical magnetic entropy change spikes. This effect usually occur if i) the material is not properly prepared before each measurement (i.e. the memory of the sample is not properly erased) or ii) if the temperature interval between adjacent isotherms is lower than the extent of hysteresis. With the aim to suppress this feature, magnetization vs. temperature of the sample can be recorded at different constant magnetic fields and $\Delta S_M(T)$ can be calculated from this data set avoiding artifacts \cite{27, 28}. Since ZFC/FC protocol meets the demand, we employed $M^{FC}$ and also $M^{ZFC}$ data for $\Delta S_M(T)$ determination in the region of applied fields up to 1 T, where thermal hysteresis in the studied Co/Au system was observed.

The relationship between the change in magnetization and entropy can be expressed by Maxwell relation \cite{29} $\mu_0(\partial M/ \partial T)_{H} = (\partial S/\partial H)_T$, which for an isothermal-isobaric process after numerical approximated integration yields \cite{30}

\begin{equation}
\begin{aligned}
\label{numericka_aproximacia}
\Delta S_M \left(\frac{T_{n+1}+T_{n}}{2}, H_{i_{max}}\right) = \\ \mu_0 \sum_{i=1}^{i_{max}}\frac{\left(M_{n+1}-M_{n}\right)}{T_{n+1}-T_{n}}\left(H_{i+1}-H_{i}\right),
\end{aligned}
\end{equation}

\noindent where $M_n$ and $M_{n+1}$ are the magnetization values measured in magnetic field $H_i$ at temperatures $T_n$ and $T_{n+1}$, respectively.

\begin{figure}
	\graphicspath{ {figures/} }
	\includegraphics[scale = 0.3]{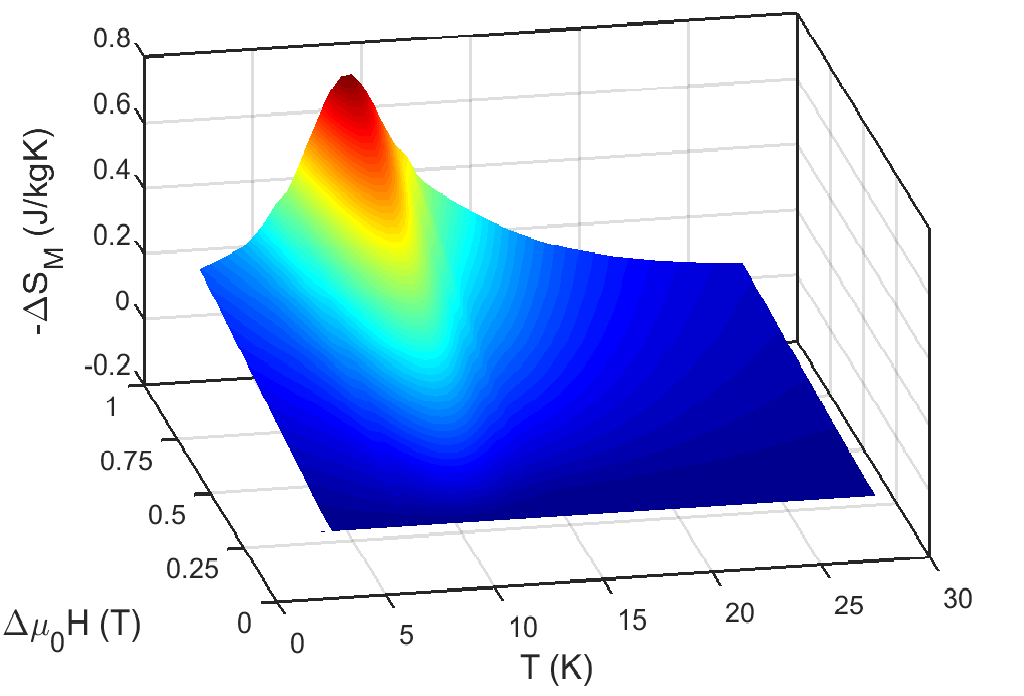}
	\caption{Magnetic entropy change dependance on temperature of Co/Au system calculated from experimental $M^{FC}$ data for magnetic field variations up to 1 T.}
	\label{figure3}
\end{figure}

Fig. \ref{figure3} presents magnetic entropy change vs. temperature dependences calculated from the experimental $M^{FC}$ data. In accordance with Maxwell relation, the slope of $M(T)$ curves determines the quantity and the sign of magnetic entropy change, while the peaks of $\Delta S_M(T)$ the occurrence of inflection points of $M^{FC}$ or $M^{ZFC}$ data. These inflection points are present in all right shoulders ($T>T_p$) of $M^{ZFC}$ curves and they are identical to the $M^{FC}$ ones, because the irreversibility does not emerge down to vicinity of $T_p$. However, in the case of left shoulders ($T<T_p$), inflection points of $M^{ZFC}$ follow the shift of $T_p$ towards lower temperatures with applied field increment and for the highest fields (100 mT, 1 T) do not even occur. Moreover, positive slope of $M^{ZFC}$ below $T_p$ could reflect the presence of inverse MCE in Co/Au nanoparticle system. Such behavior has been observed and described by Gorsse et al. \cite{31} in Tb$_{60}$Ni$_{30}$Al$_{10}$ amorphous alloy exhibiting spin glass behavior and its examination is beyond the scope of this study.

Since MCE refers to the reversible temperature change under the application or removal of an external magnetic field and $M^{ZFC}$ data are not reversible under magnetic field cycling treatment, we utilized $M^{FC}$ data for $\Delta S_M(T)$ estimation, Fig. \ref{figure3}. As expected, processed data were found in significant agreement with results established from $M^{ZFC}$ curves above $T_p$ (curves not shown), where the positions of corresponding $\Delta S_M(T)$ peaks, $T_C \sim 9$ K, along with their magnitudes were not or only slightly altered. At this point, it is worth to emphasize good correlation of $\Delta S_M(T)$ peaks' positions with freezing temperature of Co/Au nanoparticles' superspins $T_{SSG} \sim 7$ K established for the system from AC susceptibility measurements in our previous work \cite{21}. In addition, evaluation of $M(H)$ data obtained for high applied field region confirmed the presence of $\Delta S_M(T)$ peak at $T = 7.5$ K, thus at the very proximity of $T_{SSG}$ and close to $T_C$.

Finally, we compared our results with the values reported by other authors for selected nanoparticle systems. The magnitudes of magnetic entropy change at corresponding critical temperatures $T_C$ in applied field of 1 T are listed in Table \ref{table1}.

\begin{table} [h]
	\caption{\label{table1}
		Magnetic entropy change magnitudes $|\Delta S_M^{pk}|$, critical temperatures ($T_C$) and refrigerant capacities (RC) of selected nanoparticle systems in applied field 1 T.}
	\begin{ruledtabular}
		\begin{tabular}{cccc}
			
			&$T_C$&$|\Delta S_M^{pk}|$&$RC$\\
			&(K)&(J/kgK)&(J/kg)\\ \hline
			Co/Au nanoparticles\footnote{This study.}&9&0.7&4.49\\
			Co nanoparticles\footnote{Polycrystalline nanoparticles (50 nm) with average crystalline size 4 nm \cite{8}.}&13&0.91&-\\
			Co/Ag nanoparticles\footnote{Nanoparticles with polycrystalline Co core of diameter 40 nm and Ag shell of thickness 28 nm \cite{8}.}&13&0.89&-\\
			La$_{0.8}$Ca$_{0.2}$MnO$_3$\footnote{Nanoparticles prepared by sol-gel method. Average diameter is referenced in the brackets \cite{7,9}.} (17 nm) &90&0.1&20\\
			La$_{0.8}$Ca$_{0.2}$MnO$_3$ (28 nm) &210&1&75\\
			La$_{0.8}$Ca$_{0.2}$MnO$_3$ (43 nm) &236&4.2&40\\
			La$_{2/3}$Ca$_{1/3}$MnO$_3$ (60 nm) &260&0.5&-\\
			La$_{2/3}$Ca$_{1/3}$MnO$_3$ (500 nm) &260&5&-\\
			CoFe$╝_2$O$_4$\footnote{Strongly interacting non-monodisperse nanoparticles of average diameter 5 nm \cite{32}.}&223&2.5$\times10^{-2}$&-\\
		\end{tabular}
	\end{ruledtabular}
	
\end{table}

\subsection{Scaling analysis}

Numerous studies \cite{11, 14, 19, 20} showed that the occurrence of $\Delta S_M(T)$  peak ($\Delta S_M^{pk}(T)$) at peak temperature $T^{pk}$ can indicate magnetic phase transition in a system and analysis of the peak features can provide us with information on the character of possible phase transition \cite{15},  presence of minor phases in material \cite{17} or system critical behavior \cite{14}. In this study, we focused on the evaluation of $-\Delta S_M(T)$ peaks at the temperature $T>T_p$ and investigation of the characteristics of superparamagnetic to superspin glass state transition in the Co/Au nanoparticle system. With this objective we applied $\Delta S_M(T)$ scaling analysis proposed and experimentally verified for the systems undergoing the second order phase transition (SOPT) in bulk materials \cite{14}.

According to Franco et al. \cite{14} $\Delta S_M(T,H)$ curves obtained for SOPT material can be rescaled onto a single curve which does not depend on field and in which the different temperature dependencies are unified. The first step in constructing process of such master curve stems in identifying reference temperature, $T_r$, for each $\Delta S_M(T)$ dependence. Reference temperature corresponds to a certain value of magnetic entropy change and it is identified as $\Delta S_M(T_r)= \varepsilon \Delta S_M^{pk}$, where $\varepsilon$ is arbitrary selected real number between 0 and 1. The value of $\varepsilon = 0.5$ was set for this study as it is the value commonly used for the purpose of scaling analysis \cite{17, 19, 20}. Afterwards, $\Delta S_M(T)$ curves are normalized with respect to their maximum and finally the temperature axis is rescaled according to equation \cite{17}

\begin{equation}
\begin{aligned}
\label{Theta}
\theta=\begin{cases}
-(T-T^{pk})/(T_{r1}-T^{pk});T\leq T^{pk}\\
(T-T^{pk})/(T_{r2}-T^{pk});T> T^{pk}\\
\end{cases},
\end{aligned}
\end{equation}

\noindent where $T_{r1}$ and $T_{r2}$  correspond to the reference temperatures below and above the $T^{pk}$, respectively.

\begin{figure}
	\graphicspath{ {figures/} }
	\includegraphics[scale = 0.62]{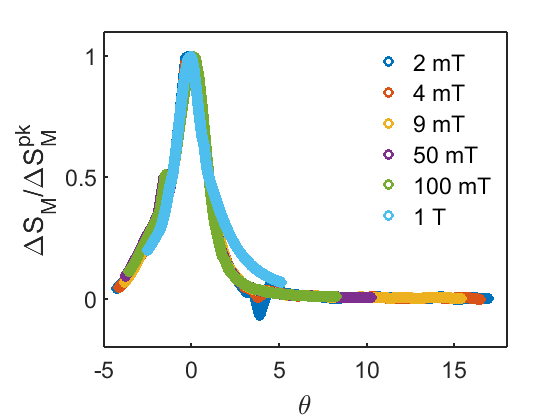}
	\caption{The colapse of $\Delta S_M(T,H)$ dependences onto single universal curve.}
	\label{figure4}
\end{figure}

\begin{figure}
	\graphicspath{ {figures/} }
	\includegraphics[scale = 0.38]{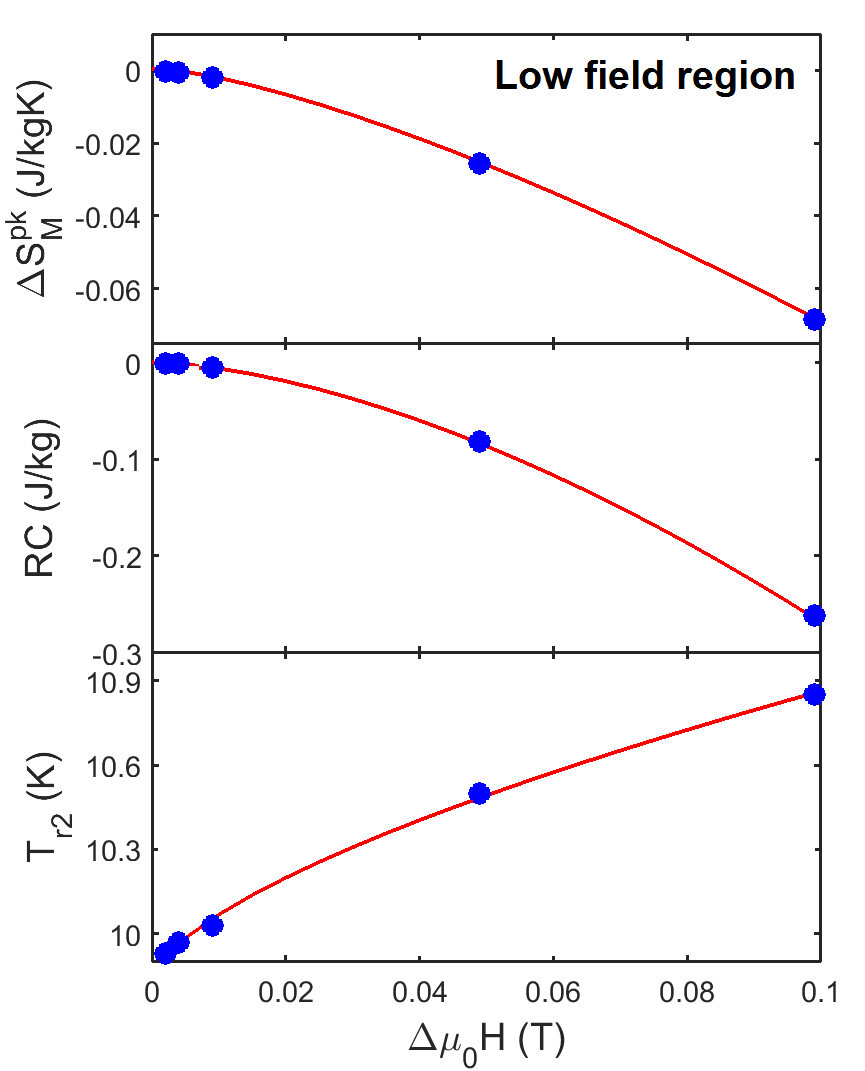}
	\caption{Low field dependence of the peak magnetic entropy change, refrigerant capacity and reference temperature ($T_{r2}$) of Co/Au nanoparticle system.}
	\label{figure5}
\end{figure}

Fig. \ref{figure4} shows the collapse of $\Delta S_M(T,H)$ data (presented in the Fig. \ref{figure3}) onto universal curve constructed for the $-\Delta S_M(T)$ maxima after employing scaling analysis according to Eq. \ref{Theta}. Analogous behavior has already been reported \cite{20} in similar Co/Ag and Ni/Ag core/shell nanoparticle systems, where the surface spin freezing transition as typical order-disorder transition was found as an origin of this behavior. Due to the existence of master curve, superspin glass freezing occurring in our Co/Au system resembles the features of the second order phase transition in the vicinity of $T_p$. The absence of thermal hysteresis in the ZFC/FC curves for fields above 1 T indicated a different nature of the transition, as can be observed in the absence of collapse of the rescaled 1 T curve onto the master curve for lower fields. In order to examine the phenomenon in more detail, further analysis of relevant magnetocaloric parameters dependence on the magnitude of applied magnetic field change has been carried out. As it was reported \cite{14}, field dependence of $\Delta S_M^{pk}$, $T_r$ and refrigeration capacity (RC) can be utilized for the calculation of critical exponents which control magnetic phase transition at the proximity of critical temperature. Refrigeration capacity is the measure of the energy that can be transferred between the hot and cold reservoirs and is defined as \cite{11}

\begin{equation}
\begin{aligned}
\label{RC}
RC(\Delta H) = \int_{T_{cold}}^{T_{hot}} \Delta S_M(T, \Delta H)dT,
\end{aligned}
\end{equation}

\noindent where $\Delta H$ is the difference between minimum and maximum applied fields. Usually, it is calculated as full width at half maximum of the  $\Delta S_M(T)$  peak times the peak value $\Delta S_M^{pk}$ and in the case of SOPT, RC should obey scaling law \cite{14} RC$\propto H^m$, $m=1+1/\delta$. Fig. \ref{figure5} shows $RC(\Delta H$) dependence up to $\Delta H$ = 100 mT of investigated Co/Au nanoparticle system and fit of experimental data yielded the value of $m = 1.68$ from which $\delta = 1.47$ was calculated. It is necessary to emphasize that we excluded the values of $\Delta S_M^{pk}$, $T_r$ and RC obtained for the field change of 1 T from corresponding data fits. This was due to the absence of thermal hysteresis in $M^{ZFC}$/$M^{FC}$ curves at $H= 1$ T, what indicates different behavior of the system in low and high applied magnetic fields. As we mentioned before, Franco et al. \cite{20} studied similar systems of pure Co, Co/Ag and Ni/Ag core/shell nanoparticles where surface spin freezing process at low temperatures was revealed. They established $m = 1.47$ from  $RC$ vs. $\Delta H$ dependence for field changes above 1 T and found out that this exponent did not vary with the change of core or shell material what points to the fact that the nature of the spin glass phase transition is not altered by material characteristics. Since our value $m = 1.68$ differs from the one determined for surface spin freezing, we suppose that transition between superparamagnetic and superspin glass state \cite{33} at low applied fields is of distinct nature. Furthermore, the comparison of $\delta = 1.47$ determined for our Co/Au system with the value of $\delta = 3$ resulting from the mean-field theory \cite{14} points to the fact that this theory is not suitable for the description of investigated SSG transition.

\begin{figure} [t]
	\graphicspath{ {figures/} }
	\includegraphics[scale = 0.57]{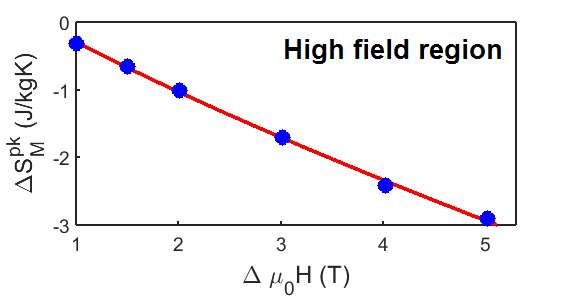}
	\caption{High field dependence of the peak magnetic entropy change of Co/Au nanoparticle system.}
	\label{figure6}
\end{figure}

In the case of materials which do not follow a mean-field approach, the magnetic equation of state of such a material in the proximity of the transition temperature can be approximately described by the Arrott-Noakes equation of state \cite{17, 34}

\begin{equation}
\begin{aligned}
\label{state}
H^{1/\gamma} =a(T-T_C)M^{1/\gamma}+bM^{1/\beta+1/\gamma},
\end{aligned}
\end{equation}

\noindent where $a$, $b$ are parameters of the equation, $\gamma$ and $\beta$ are the crittical exponents. $T_C$ is the critical temperature represented by Curie temperature in standard SOPT. In our case, we attributed critical temperature $T_C$ to superspin glass freezing temperature $T_{SSG} \sim 7$ K determined from magnetic AC susceptibility data analysis \cite{21}. If we express field dependence as $\Delta S_M \propto H^n$ and consider the ralationship  $\beta + \gamma =\beta\delta$, Eq. (\ref{state}) leads to relationship between the exponent $n$ at the Curie point and the critical exponents of the material \cite{35}

\begin{equation}
\begin{aligned}
\label{indexy}
n=1+\frac{1}{\delta}\left(1-\frac{1}{\beta}\right).
\end{aligned}
\end{equation}

Fit of experimental $\Delta S_M^{pk}(\Delta H)$ data of investigated Co/Au nanoparticle system presented in Fig. \ref{figure5} released the value of $n = 1.29$ and subsequently $\beta = 1.73$ and  $\gamma =0.81$ were calculated. Obtained values are in high discordance with the values characteristic of the mean field theory or Ising models, Table \ref{tableA}. The only critical exponent find in line with mean-field approach was $\Delta = 0.65$ established from the $T_{r2} \propto H^{1/ \Delta}$ dependence \cite{14}.
Since no thermal hysteresis was observed in the Co/Au nanoparticle system for high applied fields (above 1 T), isothermal $M(H)$ data were processed employing Eq. (\ref{numericka_aproximacia}) for the evaluation of the systems' magnetocaloric properties in high field region. Magnetic entropy change exhibited sharp peak at 7.5 K and its magnitude scaled with the applied field following power law $\Delta S_M^{pk}(\Delta H)\propto H^{0.81}$, where the exponent was obtained by fitting the experimental data, Fig. \ref{figure6}. Apparently, magnetocaloric response of the system qualitatively differs depending on low or high applied field magnitude. Due to very sharp $\Delta S_M$ peak character, insufficient experimental data hampered further scaling analysis as well as critical exponents calculations. Nevertheless, our analysis of MCE revealed distinct behavior of Co/Au nanoparticle system in low and high applied field regions and confirmed that examined transition from SPM to SSG state in low field region is of the second order type.

\begin{table} [h]
	\caption{\label{tableA}
		Determined critical exponents.}
	\begin{ruledtabular}
		\begin{tabular}{cccc}

			&$\beta$&$\gamma$ &$\delta$\\ \hline
			Co/Au nanoparticles\footnote{Exponents calculated for low applied magnetic fields.}&1.73&0.81&1.47\\
			Mean filed theory\cite{14}&0.5&1&3 \\
			3-d Ising model \cite{36} &0.31&1.25&4.82\\
			2-d Ising model \cite{36} &0.13&1.75&15\\
		\end{tabular}
	\end{ruledtabular}
	
\end{table}

\section{Conclusion}

Magnetocaloric effect has been observed in Co/Au nanoparticle system exhibiting transition from superparamagnetic to superspin glass state. Decreasing the temperature below sufficiently low value, strong inter-particle interactions become dominant inducing collective behavior of nanoparticles' magnetic moments what results in the superspin frustration and non-equilibrium dynamics. Since magnetic characteristics of SSG state differ significantly from SPM state, abrupt change in magnetic entropy of the system was expected and experimentally confirmed. The values of $|\Delta S_M^{pk}|=0.7$ J/kgK and corresponding refrigeration capacity RC$ = 4.49$ J/kg have been established for applied field $\mu_0H = 1$ T. It has been shown that despite of evident thermal hysteresis of the system observed in low applied magnetic fields, universal curve can be constructed from  $\Delta S_M(T,H)$ data recorded in field-cooling protocol. This suggests that the transition between superparamagnetic and superspin glass state evidenced in the system is of the second order (SOPT). Further data analysis by means of scaling laws valid for for magnetic materials undergoing the second order phase transition revealed significant deviations of calculated critical exponents from the values typical of mean-field theory or Ising models what suggests that application of these approaches has limitations for examined transition description. Further, established exponents do not match with those reported for nanoparticle systems where freezing of surface spins into spin glass state proceeds. This indicates that there is a need of proposing new theory describing those processes in nanoparticle systems.

\begin{acknowledgments}
This work was supported by the Slovak Research and Development Agency under the contracts APVV-14-0073, APVV-15-0520, APVV-15-0115, and VEGA (No. 1/0377/16, No. 1/0745/17) projects of Ministry of Education of the Slovak Republic and by the ERDF EU grant under No. ITMS 26220120035. The authors (A.Z and V.Z) would like to thank DESY/HASYLAB project under No. I-20110282 EC.
\end{acknowledgments}


\bibliography{references}

\end{document}